\documentclass[twocolumn,prl,aps]{revtex4}
\usepackage{graphicx}
\newcommand{\be}{\begin{equation}}
\newcommand{\ee}{\end{equation}}
\newcommand{\ba}{\begin{eqnarray}}
\newcommand{\ea}{\end{eqnarray}}
\newcommand{\ban}{\begin{eqnarray*}}
\newcommand{\ean}{\end{eqnarray*}}

\newcommand{\ket}[1]{\mbox{$ | #1 \rangle $}}
\newcommand{\bra}[1]{\mbox{$ \langle #1 | $}}

\newcommand{\one}{\leavevmode\hbox{\small1\normalsize\kern-.33em1}}
\begin{document}

\title{Long distance entanglement swapping with photons from separated sources.}

\author{H. de Riedmatten, I. Marcikic, J.A.W. van Houwelingen, W. Tittel, H. Zbinden, and N.
Gisin}

\affiliation{Group of Applied Physics, University of Geneva,
Switzerland\\}
\date{\today}
\begin{abstract}
We report the first experimental realization of entanglement
swapping over large distances in optical fibers. Two photons
separated by more than two km of optical fibers are entangled,
although they never directly interacted. We use two pairs of
time-bin entangled qubits created in spatially separated sources
and carried by photons at telecommunication wavelengths. A partial
Bell state measurement is performed with one photon from each pair
which projects the two remaining photons, formerly independent
onto an entangled state. A visibility high enough to violate a
Bell inequality is reported, after both photons have each
travelled through 1.1 km of optical fiber.
\end{abstract}
\maketitle
\begin{figure}[h!]
\begin{center}
\includegraphics[width=8cm]{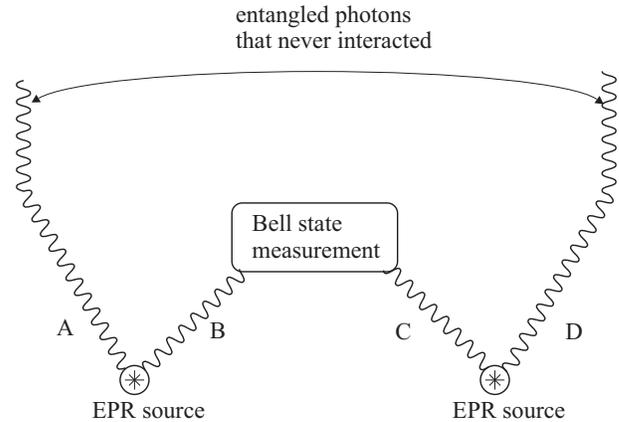}
\caption{Scheme of principle of entanglement swapping. The idea is
to start from two independent pairs of entangled particles (EPR
sources) and to subject one particle from each pair to a joint
measurement called Bell state measurement (BSM). This BSM projects
the two other particles, formerly independent onto an entangled
state \cite{zuk93} } \label{swapbasic}
\end{center}
\end{figure}

\begin{figure}[h!]
\begin{center}
\includegraphics[width=8cm]{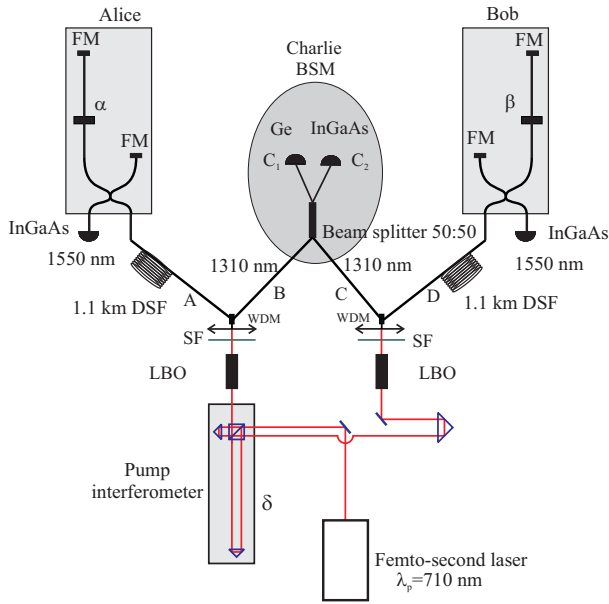}
\caption{Experimental setup. The pump laser is a mode-locked
femtosecond Ti-Sapphire laser producing 200 fs pulses at a
wavelength of 710 nm with a repetition rate of 75 MHz. After the
crystals, the pump beam are blocked with silicon filters (SF). The
Faraday mirrors (FM) are used to compensate polarization
fluctuations in the fiber interferometers} \label{setup}
\end{center}
\end{figure}

 Quantum teleportation is a process that enables the
quantum state of an object to be transferred from one place to a
distant one without ever existing anywhere in between. The quantum
teleportation channel is nothing like an ordinary channel: it
follows no path in space, but consists of entangled particles.
Entanglement is a property at the roots of quantum physics which
leads to non-local correlations between distant particles that
cannot be explained by classical physics. Entangled particles
behave as if they were a single object, non separable into its
constituents. Now, entanglement itself can be teleported, if the
state to be teleported is part of an entangled state. This
process, called entanglement swapping \cite{zuk93}, allows one
thus to concatenate quantum teleportation channels. This protocol
beautifully illustrates the oddness of quantum physics since it
enables one to entangle two particles that have never directly
interacted. Hence, two particles with no common past can act as a
single object. The principle of entanglement swapping is explained
in Fig.\ref{swapbasic}. Besides its fascinating aspect,
entanglement swapping also plays an essential role in the context
of quantum information science. It is for instance the building
block of protocols such as quantum repeaters \cite{briegel98,dlcz}
or quantum relays \cite{relay,Collins03,MarcikicJMO} proposed to
increase the maximal distance of quantum key distribution and
quantum communication. It also allows the implementation of an
heralded source of entangled photon pairs \cite{zuk93}. Finally,
it is a key element for the implementation of quantum networks
\cite{bose98}
and of Linear Optics Quantum Computing \cite{klm}. \\
The entanglement swapping protocol has been first proposed by
Zukowski and colleagues in 1993 \cite{zuk93}. The first
experimental demonstration has been reported in 1998, using
polarization entangled qubits encoded in photons around 800nm
\cite{pan98}. In 2002, an improved version of this experiment
allowed a violation of a Bell inequality with the teleported
entanglement \cite{jennewein01}, thus confirming the non-local
character of this protocol. More recently, two quantum
teleportation experiments using mode-entangled qubits have been
performed, that can be interpreted as entanglement swapping
experiments, although they differ from the original proposal since
they involved only two photons instead of four
\cite{Sciarino02,Fattal04}. Finally, an experiment demonstrating
the principle using continuous variables has also been reported
\cite{swapcontinuous}. All the experiments realized so far have
demonstrated
the principle of entanglement swapping over short distances (of the order of a meter).\\
In this paper, we present the first experimental demonstration of
entanglement swapping over large distances in optical fibers. We
use two pairs of time-bin entangled qubits encoded in photons at
telecommunication wavelengths created by parametric down
conversion (PDC). Contrary to previous swapping experiments
involving four photons, the two pairs are created in spatially
separated sources although pumped by the same laser. A partial
Bell state measurement (BSM) is performed onto one photon from
each pair, entangling the two remaining photons which have each
travelled over separated 1.1 km spools of optical fiber. A two
photon interference visibility high enough to violate a Bell
inequality is demonstrated, conditioned on a successful BSM.
Hence, two photons separated by more than two km of optical fibers
exhibit non-local correlations although they have been created in
spatially separated sources and have consequently never
interacted. The use of time-bin encoding at telecommunication
wavelength is an important extension compared to previous
experiments, since it has proven to be well suited for long
distance transmission in optical fibers \cite{marcikic04}. In
addition, time-bin entanglement can be easily extended to
high-dimensional Hilbert spaces in a scalable way with only two
photons \cite{hdr04}. Moreover, as explained below, the present
scheme is intrinsically robust against phase fluctuations and pump
laser wavelength drifts in the preparation stage, provided that we
restrain ourself to a partial BSM. Hence, this experiment can also
be considered as a (post-selected) heralded source of entangled
photons pairs robust against phase fluctuation in the preparation
stage \cite{note1}.
\\Time-bin entangled qubits can be seen as photon pairs created
in a coherent superposition of two emission times with a well
defined relative phase \cite{brendel99}. They are created first by
splitting a laser pulse into two subsequent pulses using an
unbalanced interferometer called pump interferometer. One photon
pair is then created by PDC. The down-converted photons originate
from the two pulses with a relative phase $\delta$, hence the
photon pair quantum state is  $
\ket{\phi^{+}(\delta)}=c_0\ket{0,0}+e^{i\delta}c_1\ket{1,1}$,
where \ket{0,0} corresponds to a photon pair created in the early
time-bin and \ket{1,1} to a photon pair created in the delayed
time-bin, with
$c_0^2+c_1^2=1$.\\
In our experiment, we employ two spatially separated sources of
entangled photons. In one of these sources, we create a state
$\ket{\phi^{+}(\delta)}_{A,B}$ while in the other one we create a
state
$\ket{\phi^{-}(\delta)}_{C,D}=\ket{\phi^{+}(\delta+\pi)}_{C,D}$.
Initially, the two photon pairs are independent and the total
state can be written as the tensor product: \ba
\ket{\Psi_{ABCD}}=\ket{\phi^{+}(\delta)}_{AB}\otimes\ket{\phi^{-}(\delta)}_{CD}\ea
This state can be rewritten in the form: \ba \ket{\Psi_{ABCD}}=
\frac{1}{2}[\ket{\phi^{+}}_{BC}\otimes\ket{\phi^{-}(2\delta)}_{AD}\nonumber\\+
\ket{\phi^{-}}_{BC}\otimes\ket{\phi^{+}(2\delta)}_{AD}\nonumber\\+
\ket{\psi^{+}}_{BC}\otimes
e^{i\delta}\ket{\psi^{-}}_{AD}\nonumber\\+
\ket{\psi^{-}}_{BC}\otimes e^{i\delta}\ket{\psi^{+}}_{AD}]\ea
where the four Bell states are: \ba \ket{\phi^{\pm}(\delta)}=
\frac{1}{\sqrt{2}}\left(\ket{0,0}\pm
e^{i\delta}\ket{1,1}\right) \nonumber\\
\ket{\psi^{(\pm)}}=\frac{1}{\sqrt{2}}\left(\ket{1,0}\pm
\ket{0,1}\right)\label{bellbasis}\ea When photons $B$ and $C$ are
measured in the Bell basis (Eq. \ref{bellbasis}), i.e. projected
onto one of the four Bell states via a so-called Bell state
measurement, photons $A$ and $D$ are projected onto the
corresponding entangled state. Note that when photons B and C are
projected onto the state $\ket{\psi^{+}}$ or $\ket{\psi^{-}}$, the
state of photons $A$ and $D$ is independent of the phase $\delta$
which appears only as a global factor. This means that in this
case, the creation process is robust against phase fluctuations in
the pump interferometer \cite{note2}. If however photons $B$ and
$C$ are projected onto the state $\ket{\phi^{+}}$ or
$\ket{\phi^{-}}$ the state of photons $A$ and $D$ depends on twice
the phase $\delta$. In our experiment, we make a partial BSM,
looking only at projections of photons $B$ and $C$ onto the
$\ket{\psi^{-}}$ Bell state. Apart from robustness, another
interesting feature to note is that all the four Bell states are
involved in the experiment, since we start from $\ket{\phi^{+}}$
and $\ket{\phi^{-}}$ states, make a projection onto the
$\ket{\psi^{-}}$ state, which projects the two remaining photons
onto the $\ket{\psi^{+}}$ state.\\
A scheme of our experiment is presented in Fig.\ref{setup}.
Femtosecond  pump pulses are sent to an unbalanced bulk Michelson
interferometer with a travel time difference of $\tau=1.2 ns$.
Thanks to the use of retroreflectors, we can utilize both outputs
of the interferometer, which are directed to spatially separated
Lithium triborate (LBO) non linear crystals. Collinear non
degenerate time-bin entangled photons at telecommunication
wavelength (1310 and 1550 nm) are eventually created by parametric
down-conversion (PDC) in each crystal. Because of the phase
acquired at the beam splitter in the pump interferometer there is
an additional relative phase of $\pi$ between the terms
$\ket{0,0}$ and $\ket{1,1}$ in the second output of the
interferometer. This explains why a state $\ket{\phi^+(\delta)}$
is created in one
crystal while a state $\ket{\phi^-(\delta)}$ is created in the other one.\\
 The created photons are coupled into single-mode optical
fibers and separated with a wavelength-division multiplexer (WDM).
The two photons at 1310 nm ($B$ and $C$) are sent to a Bell state
analyzer (BSA), where only the Bell state $\ket{\psi^{(-)}}_{BC}$
is detected. We use an interferometric BSA based on a standard
50-50 fiber beam-splitter \cite{weinfurter94}. It can be shown
that whenever photons $B$ and $C$ are detected in different
outputs modes and different time-bins, the desired projection is
achieved \cite{MarcikicJMO}. For this kind of measurement, the two
incoming photons must be completely indistinguishable in their
spatial, temporal, spectral and polarization mode. The
indistinguishability is verified by a Hong-Ou-Mandel experiment
\cite{hong87,hdr03}. The two photons at 1310 nm are filtered with
5 nm bandwidth interference filters (IF) in order to increase
their coherence length to 500 fs, larger than the pump pulses
duration (200 fs), which is necessary in order to make the photon
temporally indistinguishable \cite{zukowski95}. As a consequence
of the use of femtosecond pulses, the bandwidth of the
down-converted photons is large, leading to severe depolarization
effects in long fibers. Time-bin encoding is thus an advantage in
this context, since it it not sensitive to depolarization effects.
 \\
The two photons at 1550 nm, filtered to 18 nm bandwith ($A$ and
$D$) each travel over 1.1 km of dispersion shifted fiber (DSF).
Their entanglement is then analyzed with two fibre Michelson
interferometers with the same travel time difference as the pump
one. The phase of each interferometer can be varied with a piezo
actuator (PZA). In order to control the phase and to obtain a long
term stability, the fiber inteferometers are actively controlled
using a frequency stabilized laser (Dicos) and a feed-back loop on
the PZA. The phase of each fiber interferometer is probed
periodically and is locked to a user-defined value
\cite{marcikic04}. This technique allowed us to obtain a excellent
stability tested over up to 96 hours. Note that the
phase of the pump interferometer is not actively stabilized.\\
The photons are detected with avalanche photodiode (APD) single
photon detectors. One of the 1310 nm photon (detector $C_1$) is
detected with a liquid Nitrogen cooled Ge APD (NEC), with an
efficiency of around 10$\%$ for 40KHz of dark counts. The three
other photons are detected with InGaAs APDs (id-Quantique) with an
efficiency of $30\%$ for a dark count probability of around
$10^{-4}$ per ns. The trigger signal for those detectors is given
by a coincidence between the Ge APD and the emission time of the
pump pulses. The coincidence events between different detectors
are recorded with a multistop time-to digital converter (TDC). The
coincidence between the Ge APD and the emission time of the laser
is used as START while the other APDs are used as STOPs. Note that
the classical information about the BSM is delayed electronically
by roughly 5 $\mu s$, corresponding to the travel time of the 1550
nm photons inside optical fibers. Hence, the swapping process is
completed when the photons are already two km apart. A home made
programme allows us to register any desired combination of
coincidence count rate between the four detectors, which is useful
to characterize the stability of the whole setup during the
measurement process. In our experiment, the average pump power
 for each source was about 80 mW, leading to a probability
 of creating an entangled pair per laser pulse of around $6\%$.\\
If entanglement swapping is successful, the two photons $A$ and
$D$ at 1550 nm should be in the entangled state
$\ket{\psi^{(+)}}$, conditioned on a projection on the
$\ket{\psi^{(-)}}$ Bell state. However, as real measurements are
imperfect, there will be some noise, that we suppose equally
distributed between all possible outcomes. Hence, the created
state can be written: \ba \rho=F_2\ket{\psi^{(+)}}\bra{\psi^{(+)}}
+\frac{1-F_2}{3}(\ket{\psi^{(-)}}\bra{\psi^{(-)}}\nonumber\\
+\ket{\phi^{(+)}}\bra{\phi^{(+)}}
+\ket{\phi^{(-)}}\bra{\phi^{(-)}}) \nonumber\\=V
\ket{\psi^{(+)}}\bra{\psi^{(+)}} + \frac{(1-V)}{4}\one \ea where
$V$ is the visibility and $F_2$ the two-qubit fidelity defined as:
$F_2=\bra{\psi^{(+)}}\rho\ket{\psi^{(+)}}$. V is related to $F_2$
as: \ba V=\frac{4F_2-1}{3}\ea The Peres criteria \cite{peres96}
shows that the two photons are entangled (i.e. in a non separable
state) if $V> \frac{1}{3}$, and consequently if $F_2>
\frac{1}{2}$. It can also be shown that Bell's inequalities can be
violated if $V> \frac{1}{\sqrt{2}}$ \cite{clauser}.
\\
To verify the entanglement swapping process we perform a
two-photon interference experiment with the two photons at 1550
nm, conditioned on a successful BSM. This is done by sending the
two photons $A$ and $D$ to two interferometers. The evolution of
$\ket{\psi^{(+)}}$ in the interferometers is: \ba
\ket{\psi^{(+)}}\rightarrow
\ket{0_A,1_D}+e^{i\alpha}\ket{1_A,1_D}+e^{i\beta}\ket{0_A,2_D}\nonumber\\
+e^{i(\alpha+\beta)}\ket{1_A,2_D}+\ket{1_A,0_D}+e^{i\alpha}\ket{2_A,0_D}\nonumber\\
+e^{i\beta}\ket{1_A,1_D}+e^{i(\alpha+\beta})\ket{2_A,1_D}
\label{evol}\ea where $\ket{i_A,j_D}$ corresponds to an event
where the photon $A$ is in time-bin $i$ and the photon $D$ is in
time-bin $j$. A photon travelling through the long arm of an
interferometer passes from time-bin $i$ to time-bin $i+1$. If the
arrival time difference between photon $A$ and $D$ are recorded,
Eq. (\ref{evol}) shows that there are 5 different time windows,
with $\Delta \tau =t_A-t_B=\{0,\pm \tau,\pm2\tau\}$. This is in
contrast with previous experiments using time-bin entangled qubits
in the state $\ket{\phi^{(\pm)}}$, where only three time-windows
were present (see e.g.\cite{marcikic04}). If only the event with
$\Delta \tau =0$ are selected, there are two indistinguishable
events leading to a coincident count rate: \ba R_c \sim
1+Vcos(\alpha -\beta)\ea where $V$ is the visibility of the
interference which can in principle attain the value of 1 but is
in practice lower than 1 due to various experimental
imperfections.
\begin{figure}[h!]
\begin{center}
\includegraphics[width=8cm]{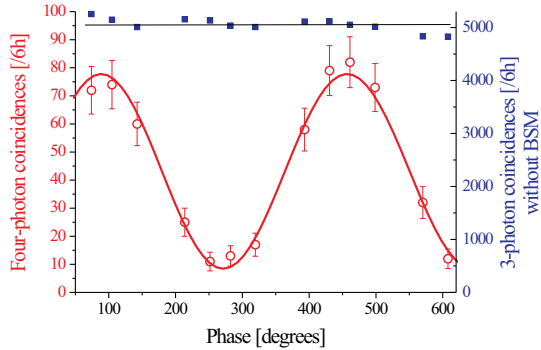}
\caption{Two-photon interferences for swapped photons, as a
function of the phase of one interferometer. Plain squares
represents the detection between photons $A$ and $D$, without
conditioning on the BSM. Errors bars are too small to be
represented. Open circles represent four-photon coincidences, i.e
two photon interference conditioned on a BSM } \label{results}
\end{center}
\end{figure}
Fig.\ref{results} shows a measurement of two photon interference.
The plain squares represent coincidences between Alice's and Bob's
photons, without conditioning on a BSM as a function of the phase
of one interferometer. The fact that the coincidence count rate
does not vary significantly with the phase is a confirmation that
the two photons are completely independent in this case. However,
if we now condition on a successful BSM (open circles), we see a
sinusoidal variation with a fitted raw (i.e. without noise
subtraction) visibility of $(80\pm4\%)$, leading to a fidelity
$F_2$ of $(85 \pm 3.25 \%)$ high enough to demonstrate a
teleportation of entanglement and to violate a Bell inequality
with the teleported photons by more than two standard deviations.
The whole measurement lasted 78 hours, which demonstrates the
robust character of our scheme. The non perfect visibility of the
interference fringe is attributed mainly to the limited fidelity
of the BSM.  The main limiting factor is the non-vanishing
probability of creating multiple photon pairs in one laser pulse,
due to the probabilistic nature of PDC \cite{hdr03,scarani04}. The
visibility could be improved by reducing the pump power but this
would reduce the four-photon coincidence count rate. Note that the
key parameters in order to increase the four photon coincidence
count rate without degrading the correlations are the quantum
efficiencies of detectors and the coupling efficiencies into the
single mode fibers.

 In summary, we have reported the first
demonstration of entanglement swapping over long distance in
optical fibers. We used two pairs of time-bin entangled qubits
encoded into photons at telecommunications wavelengths and created
in spatially separated sources. The visibility obtained after the
swapping process was high enough to demonstrate a teleportation of
entanglement and to infer a violation of Bell inequalities with
photons separated by more than 2 km of optical fibers that have
never directly interacted. This constitutes a promising approach
to push quantum teleportation and entanglement swapping
experiments out of the lab, using the existing optical fiber
network.

 The authors would like to thank Claudio Barreiro and
Jean-Daniel Gautier for technical support. Financial support by
the Swiss NCCR Quantum Photonics, and by the European project
RamboQ is acknowledged.


\begin{references}
\bibitem{note1} Because of the probabilistic nature of PDC, the probability to create two photon pairs in one
 source is the same as the probability of creating one pair in
 each source. Hence, the four photons need to be detected in
 order to post-select the cases where (at least) one pair is
 created in each source, which are the events we are interested
 in.
\bibitem{note2} Note that this is valid only if the laser used is
the same for both sources. In case of distant sources pumped by
independent lasers, the pump interferometers phases should be
stabilized.
\end{references}
\end{document}